\documentclass[a4paper,dvips 2t,nofootinbib]{revtex4}
\usepackage{setspace}
\usepackage{epsfig,graphics,amsmath,amssymb}
\usepackage{rotate}
\usepackage{latexsym}
\usepackage{graphicx}

\newcommand{\mathsym}[1]{{}}
\newcommand{\unicode}[1]{{}}

\begin{document}

\title{\bf A new parameter space study of the fermionic cold dark matter model}
\author{Z. Bagherian} \author{ M. M. Ettefaghi}\email{mettefaghi@qom.ac.ir} \author{Z. Haghgouyan} \author{R. Moazzemi} \email{r.moazzemi@qom.ac.ir}

\affiliation{Department of Physics, University of Qom, Ghadir Blvd., Qom 371614-611, I. R. Iran}

\begin{abstract}
We consider the standard model (SM) extended by a gauge singlet fermion as cold dark matter (SFCDM) and a gauge singlet scalar (singlet Higgs) as a mediator. The parameter space of the SM is enlarged by seven new ones. We obtain the total annihilation cross section of singlet fermions to the SM particles and singlet Higgs at tree level. Regarding the relic abundance constraint obtained by WMAP observations, we study the dependency on each parameter separately, for dark matter masses up to 1 TeV. In particular, the coupling of SFCDM to singlet Higgs $g_s$, the SFCDM mass $m_\psi$, the second Higgs mass $m_{h_2}$, and the Higgs bosons mixing angel $\theta$ are investigated accurately. Three other parameters play no significant role. For a maximal mixing of Higgs bosons  or at resonances, $g_s$ is applicable for the perturbation theory at tree level.  We also obtain the scattering cross section of SFCDM off nucleons and  compare our results with experiments which have already reported data in this mass range;  XENON100, LUX, COUPP and PICASSO collaborations. Our results show that the SFCDM is excluded by these experiments for choosing parameters which are consistent with perturbation theory and relic abundance constraints.
\end{abstract}
\maketitle

\section{Introduction}
There exist several evidences that indicate the most fraction of matter in the Universe is consisted of unknown particles called as Dark Matter (DM) \cite{rev1,rev2}. Namely, the observations by Wilkinson Microwave Anisotropy Probe (WMAP) for the study of fluctuations in the cosmic microwave background (CMB) radiation show that the Universe consists of $27\%$ matter and the rest $73\%$  is an unknown energy called Dark Energy \cite{Spergel}. Baryonic matter composes only less than $15\%$ of the matter content of the universe and the others remain unknown. The most important constraints and properties which need to be obeyed by DM candidates have been discussed in \cite{Taoso}. The evidences offer that the DM candidates are mostly stable, non-baryonic, massive, non relativistic and have insignificant or very weak interactions with other particles. These types of DM are often called as cold DM (CDM) or weakly interacting massive particles (WIMP). Since the particles accommodated in the standard model (SM) cannot play the DM role, it is one of the most important motivations for the extension of the SM. For instance supersymmetric models with R parity \cite{rev1,rev2}, the
extra dimensional models with conserved Kaluza-Klein (KK) parity \cite{kk}, the T-parity conserved
little Higgs model \cite{little} and so on, provide  WIMP DM candidates. Although all sectors of the SM have passed every experimental test, there does not exist any experimental confirmation for all its extensions. Therefore, some authors extend, minimally, the SM to explain anomalous problems such as DM.  For instance, it is possible to have a CDM candidate if the SM is extended by a gauge singlet scalar particle \cite{ss}. Also, a gauge singlet fermion as a CDM is investigated by \cite{sf1,sf2}, and in \cite{Queiroz} authors defined a scenario  where a singlet right-handed neutrino is added together with a charged and a neutral singlet scalars.

In particular, one can achieve a renormalizable theory for DM if the SM is extended by a singlet fermion as CDM (SFCMD) and a singlet scalar Higgs as a mediator \cite{sf2}. That is, in this theory, SFCDM interacts with SM particles through its coupling to singlet Higgs. The mixing between the SM Higgs and the singlet Higgs leads to the coupling of both Higgses  with SFCDM. The relic abundance constraint and direct detection bounds have been studied for the SFCDM masses below 100 GeV in \cite{sf2} and it has been shown that all relic abundance and LEP2 allowed regions are excluded by direct detection bounds except at resonance. The SFCDM annihilation into two photons under relic abundance constraint has been obtained and compared with Fermi-Lat bounds for masses below 200 GeV in \cite{em}. Moreover, in Ref. \cite{others} authors show that SFCDM can be such a dark matter candidate, simultaneously providing the correct thermal relic density which is consistent with relic abundance. Also in Ref. \cite{others2} SFCDM has been studied to see if it can explain the recent LHC data while fulfilling other observational and cosmological requirements.

 Recently, an updated analysis of SFCDM model has been done for masses up to 1 TeV with focusing on its direct detection prospects \cite{recent}. This analysis is based on a sample of about $10^5$ random models satisfying the usual theoretical and experimental constraints, including the dark matter bounds. Also, the electroweak phase transition has been studied for SFCDM model in \cite{phase,2014}.
 In this paper, we give an investigation of the parameter space of SFCDM model for masses up to 1 TeV, independently. Although, an annihilation cross section of SFCDM into the SM particles and two singlet Higgs bosons (with some missprints) reported in \cite{phase}, we have obtained the total cross section (including three Higgs bosons in final state).
  In particular, we study the role of every parameter which provides a viable framework for more accurate analysis in future. As we shall see, for instance, the mixing parameter between the SM Higgs and singlet Higgs is a relevant parameter and constraining it leads to significant bounds on the parameter space. We consider the relic abundance constrain to analyze  parameter space. It is noticeable that we use perturbation theory, so that those values of the coupling constants which are larger than one are unreasonable in this framework. This means that, respecting the relic density constraint we only consider points in parameter space which lead to applicable coupling constant in perturbation theory, and we do not have any judgment on the other points. In the other words, if we decide to use of perturbation theory, we naturally have to work with a set of parameters which satisfy the perturbation condition; at such a regime we compare our results with experimental data and the possible exclusion points are actually physical.
We also study the possibility of the direct detection of SFCDM by comparing the theoretical results, for those coupling constants consistent with our perturbation theory, with the recent experimental results such as XENON100 \cite{XENON100}, LUX \cite{LUX}, COUPP \cite{coupp} and PICASSO \cite{picasso}.

The paper is organized as follows: in the next section, we briefly review the extended SM by a gauge singlet fermion and a gauge singlet scalar called singlet Higgs.
In Section \ref{sec3}, the sensitivity of the annihilation cross section of SFCDM to every free parameter under relic abundance constraint is studied. In Section \ref{s3}, we obtain the cross section of the scattering of SFCDM off nucleons for two chosen sets of parameters consistent with perturbation theory, then we compare our results with the recently reported data by XENON100, LUX, COUPP and PICASSO collaborations. Finally, we give a summary and discussion in the last section.

\section{SFCDM Model}
The most minimal extension of the SM accommodating a CDM candidate is achieved by adding gauge singlet particles. In the case of scalar, one needs only a gauge singlet scalar field with zero vacuum expectation value to have a renormalizable theory for DM \cite{ss}. Otherwise, singlet fermion can also play the dark matter role (SFCDM) provided that it has a very weak interaction with the SM particles \cite{sf1,sf2}. To accommodate this by a renormalizable manner,
a singlet Higgs, in addition to the usual Higgs doublet, is needed as mediator between SFCDM and the SM particles \cite{sf2}. Therefore, the Lagrangian of SFCDM model can be decomposed as follows:
 \begin{equation} \label{sfcdm} {\cal L}_{\text{SFCDM}}={\cal L}_{\text{SM}}+{\cal L}_{\text{hid}}+{\cal L}_{\text{int}}
,\end{equation}

 where ${\cal L}_{\text{SM}}$ indicates the usual SM Lagrangian. ${\cal L}_{\text{hid}}$ is  the hidden sector Lagrangian
\begin{eqnarray} {\cal L}_{\text{hid}}={\cal L}_\psi+{\cal L}_{\text{S}}-{g_s \overline \psi  \psi S},
\end{eqnarray}
 in which apart form the last term, the interaction between the SFCDM and singlet Higgs with coupling constant $g_s$, ${\cal L}_\psi$ and ${\cal L}_{\text{S}}$ are the free Lagrangians of SFCDM
\begin{equation}{\cal L}_\psi=\bar{\psi}(i\partial\!\!\!/-m_0)\psi,\end{equation}
and singlet Higgs
\begin{equation}\label{selfS}{\cal L}_{\text{S}}=\frac 1 2 (\partial_\mu S)(\partial^\mu
S)-\frac{m_0^2}{2}S^2-\frac{\lambda_3}{3!}S^3-\frac{\lambda_4}{4!}S^4,\end{equation}
respectively. The last term of Eq.(\ref{sfcdm}), ${\cal L}_{\text{int}}$, is related to the interaction between the singlet Higgs and the SM doublet Higgs
\begin{equation}  \label{hs}
{\cal L}_{\text{int}}=-\lambda_1H^\dag HS-\lambda_2H^\dag
HS^2.
\end{equation}
Here, the coupling constants $\lambda_{1}$and$\lambda_{2}$ have one and zero mass dimension, respectively. The scalar potentials given by Eqs. (\ref{selfS}) and
(\ref{hs}) along with the scalar potential of Higgs doublet, $-\mu^2H^\dag H+\lambda_0(H^\dag
H)^2$, are minimized by
\begin{equation}
 \langle H\rangle=\frac{1}{\sqrt{2}}\left(
                             \begin{array}{c}
                               0 \\
                               v_0
                             \end{array}\right),
\end{equation}
and $\langle S\rangle=x_0$, where $v_0$ and $x_0$ are, respectively, the SM Higgs and singlet Higgs values which minimize the classical total potential.
Hence, they have to satisfy the following relations:
 \begin{eqnarray}
&&\mu^2=\lambda_0v_0^2+(\lambda_1+\lambda_2x_0)x_0,\nonumber\\
&&m_0^2=-\frac{\lambda_3}{2}x_0-\frac{\lambda_4}{6}x_0^2-\frac{\lambda_1v_0^2}{2x_0}-\lambda_2v_0^2.
\end{eqnarray}
We define the fields $h$ and $s$ as departure from the vacuum expectation values corresponding to the SM and the singlet Higgs, respectively. Therefore, after symmetry breaking $H$ and $S$ are replaced by
\begin{equation}
 H=\frac{1}{\sqrt{2}}\left(
                             \begin{array}{c}
                               0 \\
                               h+v_0
                             \end{array}\right),
\end{equation}
and
\begin{equation}
S=s+x_0.
\end{equation}
The mass matrix elements are given by
\begin{eqnarray}
&&\mu^2_h\equiv\frac{\partial^2V}{\partial
h^2}\Big{|}_{h=s=0}=2\lambda_0v_0^2,\nonumber\\
&&\mu^2_s\equiv\frac{\partial^2V}{\partial
s^2}\Big{|}_{h=s=0}=\frac{\lambda_3}{2}x_0+\frac{\lambda_4}{3}x_0^2-\frac{\lambda_1v_0^2}{2x_0},\nonumber\\
&&\mu^2_{hs}\equiv\frac{\partial^2V}{\partial h\partial
s}\Big{|}_{h=s=0}=(\lambda_1+2\lambda_2x_0)v_0.
\end{eqnarray}
Diagonalizing the mass matrix we obtain the mass eigenstates as follows:
\begin{eqnarray}
 h_1=\sin\theta s+\cos\theta h,\nonumber \\
h_2=\cos\theta s-\sin\theta h,
\end{eqnarray}
where the mixing angle $\theta$ is defined by
\begin{eqnarray} \label{mixing}\tan\theta\equiv\frac{y}{1+\sqrt{1+y^2}},
\end{eqnarray}
with $y=\frac{2\mu^2_{hs}}{(\mu^2_h-\mu^2_s)}$. The mass eigenvalues are
\begin{eqnarray}
m^2_{1,2}=\frac{\mu_h^2+\mu_s^2}{2}\pm\frac{\mu_h^2-\mu_s^2}{2}\sqrt{1+y^2}.
\end{eqnarray}
From the definition of the mixing angle (\ref{mixing}), we have $|\cos\theta|>\frac{1}{2}$.
Therefore, $h_1$ is the SM Higgs-like scalar while $h_2$ is the singlet-like one.
The singlet fermion  has mass $m_\psi=m_0+ g_Sx_0$ which is an independent parameter
in the model.
\section{Computing relic density}\label{sec3}
\begin{figure}[th]
\centerline{\includegraphics[width=15.5cm]{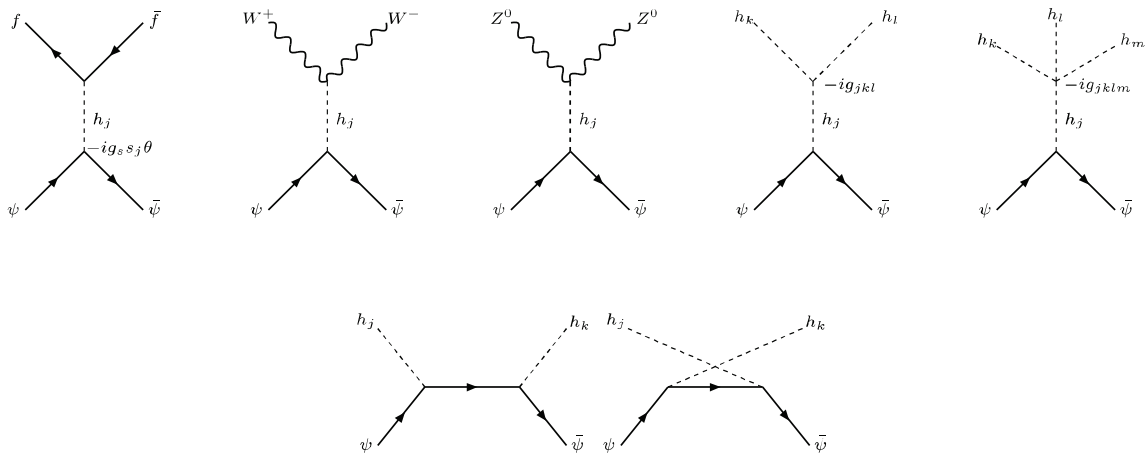}}
\caption{The Feynman diagrams for the annihilation of  singlet fermion pairs into, SM particles, two and three Higgs bosons at tree level. The vertex factor of three (four) Higgs boson lines,  $-ig_{ijk}$ ($-ig_{ijkl}$), is symmetric under permutations of their subscripts. For three Higgs bosons in final state only the dominant Feynman diagrams are shown. Obviously, the first row is due to the $s$-channel while the second row indicates the $t$- and $u$-channels.}
\label{fig1}
\end{figure}

 In the WIMP scenario, a DM particle can be  produced thermally  through a `freeze-out' mechanism. In fact, when the interaction rate of a particle species in the early universe drops below the expansion rate of the universe, it falls out of the equilibrium and its number density in the comoving volume remains constant. This  production  mechanisms arises from $\overline \psi  \psi $ pair annihilation into the SM fermions, the gauge bosons and the Higgs boson. The evolution of the density number of a singlet fermion, $n_\psi$, is given by the following Boltzmann equation:
\begin{equation}
\frac{dn_\psi}{dt}+3Hn_\psi=-\left\langle\sigma v_{\text{ann}}\right\rangle\left[n_\psi ^2-\left(n_\psi ^{\text{eq}}\right)^2\right],
\end{equation}
where $\left\langle {\sigma v_{\text{ann}} } \right\rangle$ is the thermally averaged annihilation cross section times the relative velocity, and $n_\psi ^{\text{eq}}$ indicates the equilibrium density number of $\psi$. As we said above, when the the expansion rate of the universe exceeds the interaction rate of a DM, our particle species falls out of thermodynamic equilibrium. Hence, the relic density, defined as the ratio of the present density to the critical density, $\Omega_\psi h^2$, is written roughly as follows:
\begin{equation}
\Omega_\psi h^2\approx\frac{(1.07\times10^9)x_F}{\sqrt{g_*}M_{\text{Pl}}(GeV)\left\langle\sigma v_{\text{ann}}\right\rangle},
\end{equation}
where $g_*$ is the effective degrees of freedom for the relativistic quantities in equilibrium \cite{colb}. The inverse freeze-out temperature $x_F=m/T_F$ is determined by the following iterative equation:
\begin{equation}
x_F=\ln\left(\frac{m_\psi}{2\pi^3}\sqrt{\frac{45M_{\text{Pl}}}{2g_*x_F}}\left\langle\sigma v_{\text{ann}}\right\rangle\right).
\end{equation}
We can obtain $\left\langle {\sigma v_{\text{ann} }} \right\rangle$ from \cite{gondolo}
\begin{equation}
\left\langle\sigma v_{\text{ann}}\right\rangle=\frac{1}{8m_\psi^4 TK_2^2\left(\frac{m_\psi}{T}\right)}\int_{4m_\psi^2}^\infty ds\sigma _{\text{ann}} \left( s \right)\left(s-4m_\psi^2\right)\sqrt{s}K_1\left(\frac{\sqrt{s}}{T}\right),
\end{equation}
where $K_{1,2}$ are the modified Bessel functions. In the Appendix, we obtain $\sigma_{ann}v_{rel}$ applicable for throughout mass range of singlet fermion. Note that, as the mass of DM increases new channels for annihilation to the Higgs bosons are opened in addition to the SM channels. We have listed the relevant Feynman diagrams in Fig. \ref{fig1}. These diagrams are at tree level, so we should respect the perturbation in our calculations.
\begin{figure}[th]
\centerline{\includegraphics[width=12.5cm]{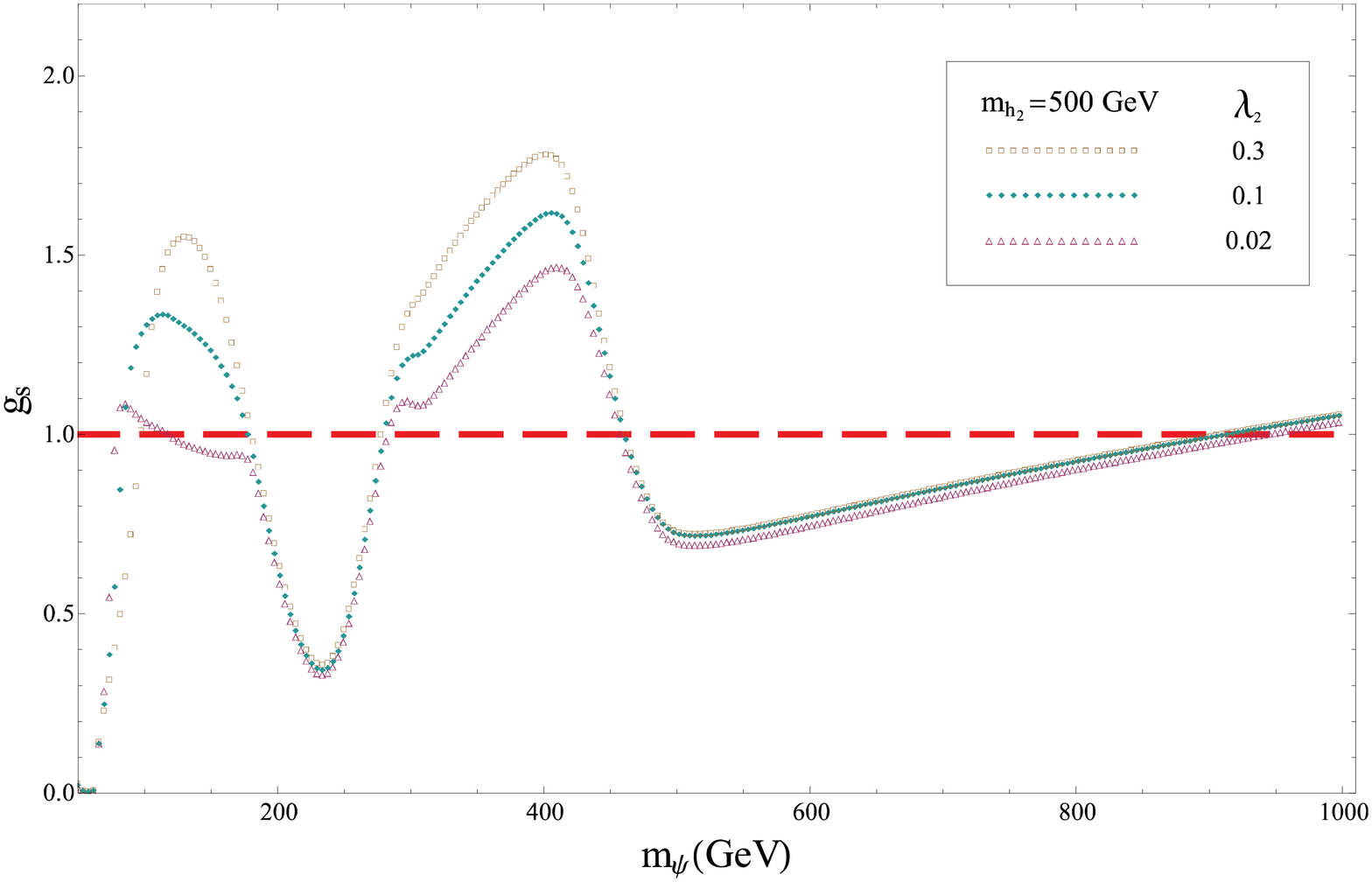}}
\caption{\small  $g_s$ vs SFCDM mass for different values of $\lambda_2$ ($\theta=0.1$ and $m_{h_2}=500$ GeV).
 }\label{f1}
\end{figure}

Now, we study the allowed parameter space consistent with the relic abundance constraint obtained by WMAP observations \cite{Spergel}. In addition to the SM parameters, here we have seven independent ones: singlet fermion mass $m_\psi$, a coupling constant between singlet fermion and singlet Higgs $g_S$, the mass of the singlet Higgs $m_0$ and its self-interaction couplings $\lambda_3$ and $\lambda_4$, and the coupling constants between singlet Higgs and SM one $\lambda_1$ and $\lambda_2$.  We encounter a new set of parameters after spontaneous symmetry breaking: $m_\psi$, $g_S$, second Higgs mass $m_{h_2}$, $\lambda_1$, $\lambda_2$, $\lambda_3$, $\lambda_4$ and  mixing angel between Higgs bosons $\theta$ which is not an independent parameter. The SM Higgs boson mass is fixed to 125 GeV according to the recent CMS and ATLAS results \cite{atlas,cms}. The VEV of our singlet Higgs, $x_0$, is completely determined by minimization of potential as follows:
 \[x_0=-\frac{1}{4v\lambda_2}\left[(m_{h_1}^2+m_{h_2}^2-4v^2\lambda_0)\tan2\theta+2v\lambda_1\right].\]

Note that the vertex factors  $g_{ijk}$ (in Fig. \ref{fig1}), corresponding to the two Higgs bosons in final state (ignoring the processes with three Higgs bosons in final state due to the kinematically suppression), come from  ${\cal L}_{\rm{int}}$ and two last terms of ${\cal L}_S$, after symmetry breaking. Therefore,  all $\lambda$'s contribute in cross section almost equally (specially around the maximal mixing). This point also can be checked from the cross section obtained in Appendix.   It depends on $\lambda_1$, $\lambda_2$, $\lambda_3$ and $\lambda_4$ via $g_{ijk}$'s and $g_{ijkl}$'s  (defined in Eq. (\ref{couplings})) roughly in similar way. We illustrate  effects of $\lambda_2$, for instance, in Fig. \ref{f1}. This figure shows $g_s$ vs DM mass by fixing the other parameters as follows: mixing angle $\theta=0.1$ and $m_{h_2}=500$ GeV, for instance, and some fixed values for the other $\lambda$'s as will be mentioned. We see that the variation of $\lambda_2$ does not significantly change the results in particular at the region where $g_s$ is smaller than 1. One can similarly check that the variation of $\lambda_1$, $\lambda_3$ and $\lambda_4$ have no significant effect too. Therefore, we fix $\lambda_1/\Lambda$, $\lambda_3/\Lambda$, $\lambda_2$ and $\lambda_4$ to value 0.1 which can be applicable in our perturbation framework. Here $\Lambda$ is a scale in our problem which we take it 100 GeV.

\begin{figure}[th]
{ (a)}\raisebox{-2cm}{\includegraphics[width=8cm]{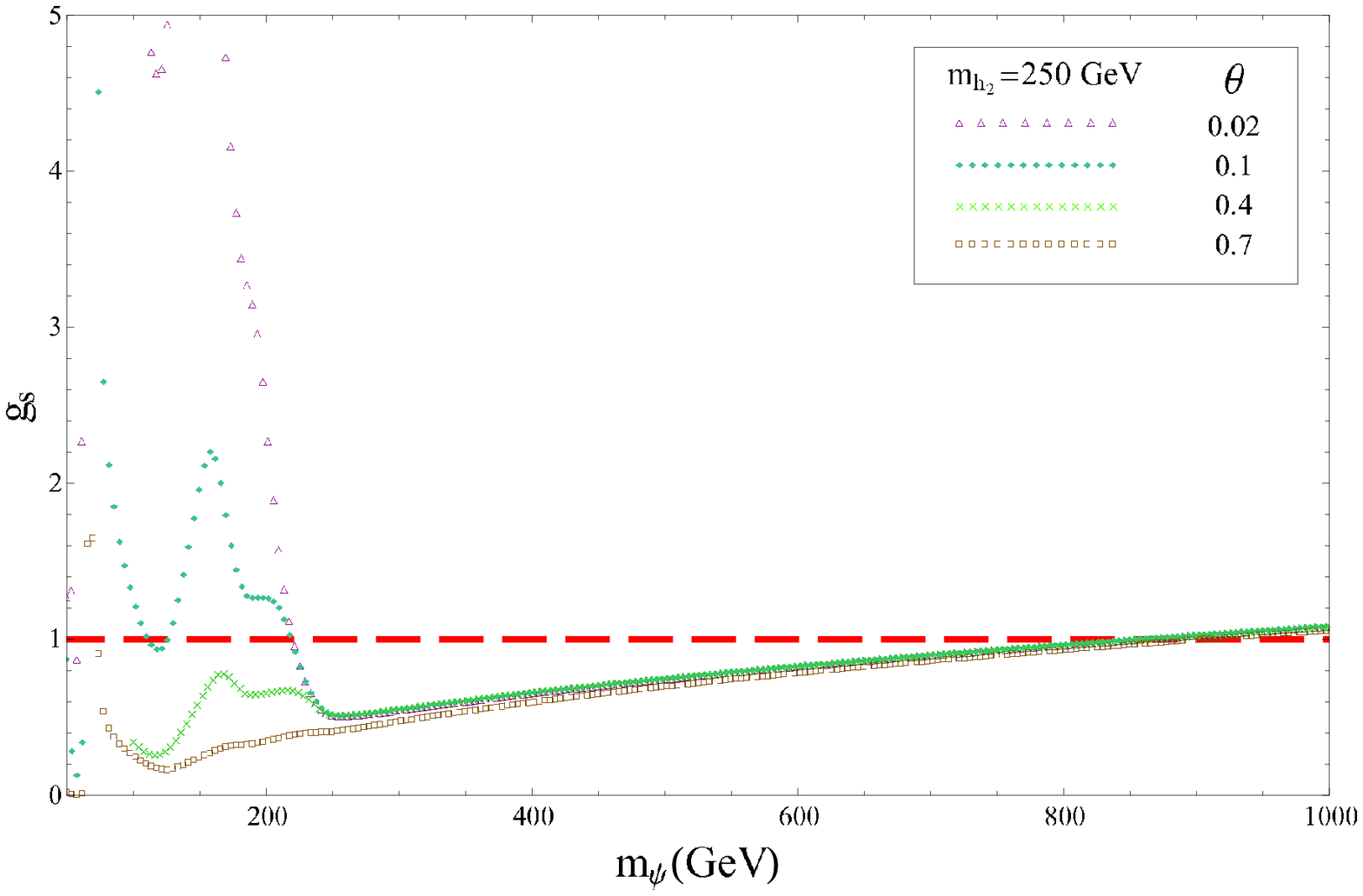}}(b)\raisebox{-2cm}{\includegraphics[width=8cm]{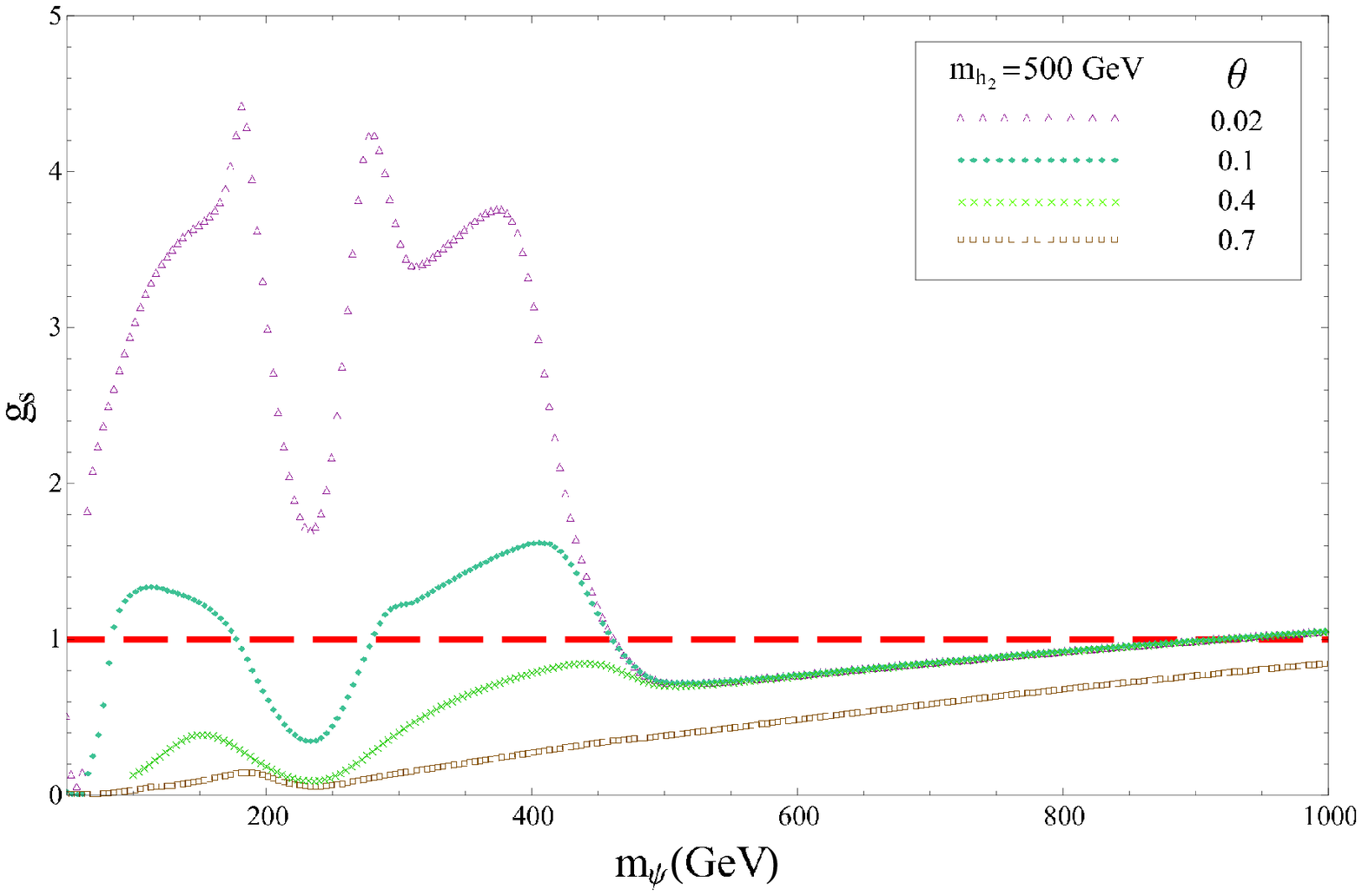}}\\
(c)\raisebox{-2cm}{\includegraphics[width=8cm]{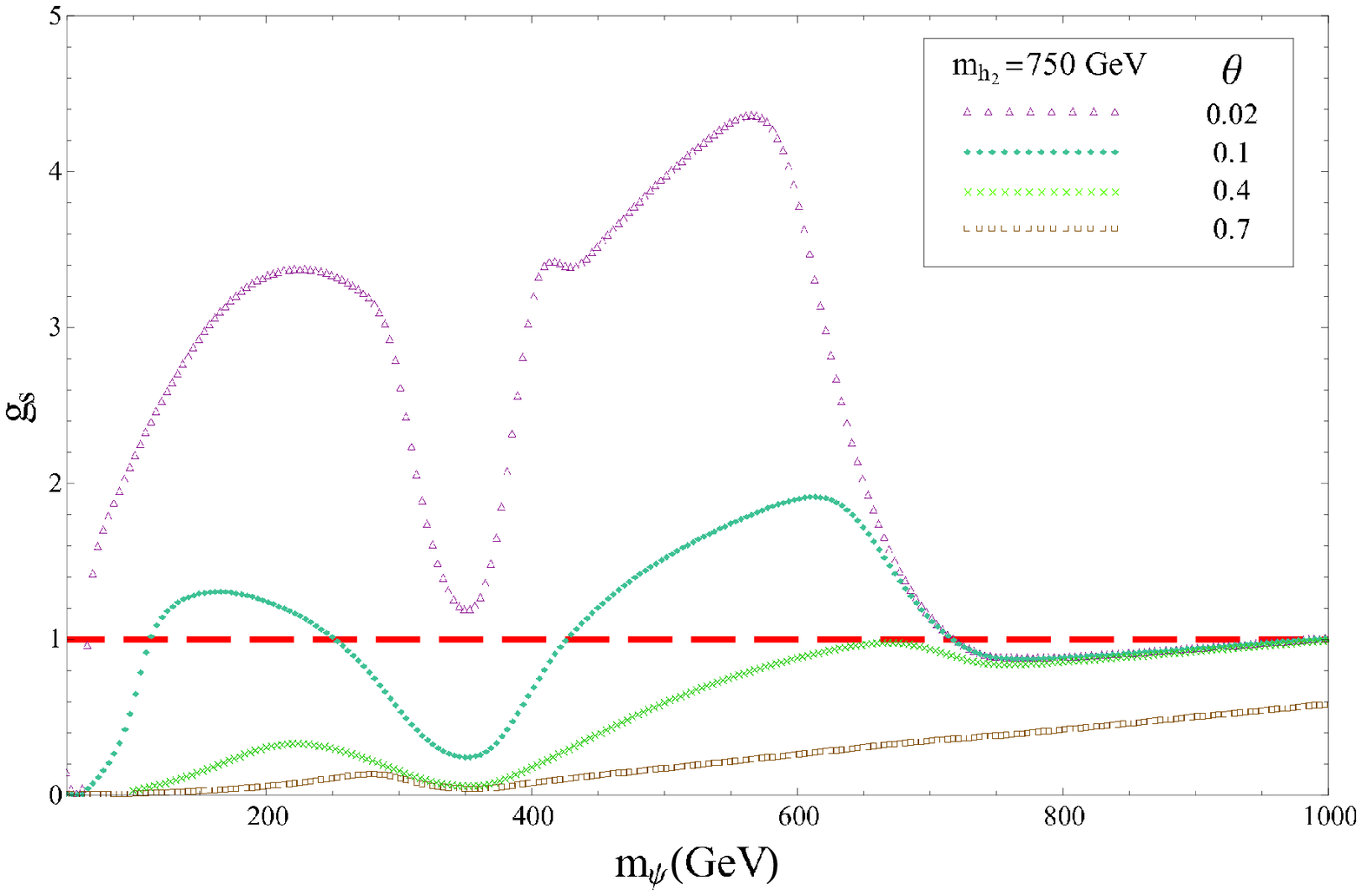}} (d)\raisebox{-2cm}{\includegraphics[width=8cm]{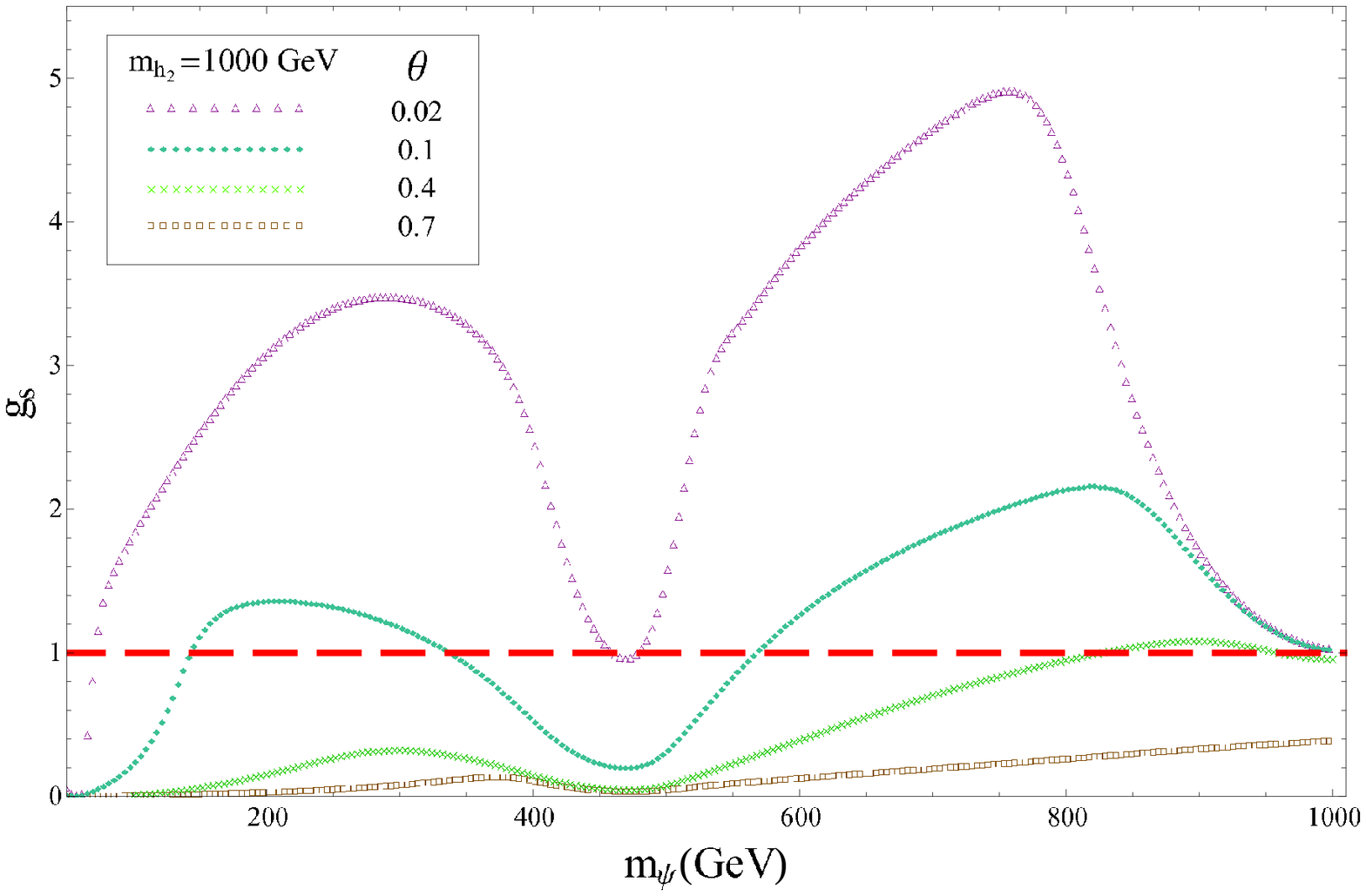}}
\caption{\small  $g_s$ vs SFCDM mass for different values of $\theta$ and $m_{h_2}$.
 }\label{f2}
\end{figure}

The other free parameters, which may have important effects, are the mixing angel $\theta$, and the mass of the second Higgs $m_{h_2}$. We try to clarify these effects in some diagrams. In Figs. \ref{f2}(a), \ref{f2}(b), \ref{f2}(c) and \ref{f2}(d)  one can see $g_s$ vs SFCDM mass for $m_{h_2}=250,500,750$ and 1000 GeV, respectively,  and different values of $\theta$ for each of which figures. These figures show that the coupling constant $g_s$ lies into perturbation regime for almost maximal mixing angle $\theta=0.7$. It is also obvious that the smaller coupling constant $g_s$ is due to the larger mixing parameter $\theta$. We see that for $\theta=0.1$ only in the resonance regions $g_s$ is smaller than 1.
\begin{figure}
{(a)\raisebox{-2cm}{\includegraphics[width=8cm]{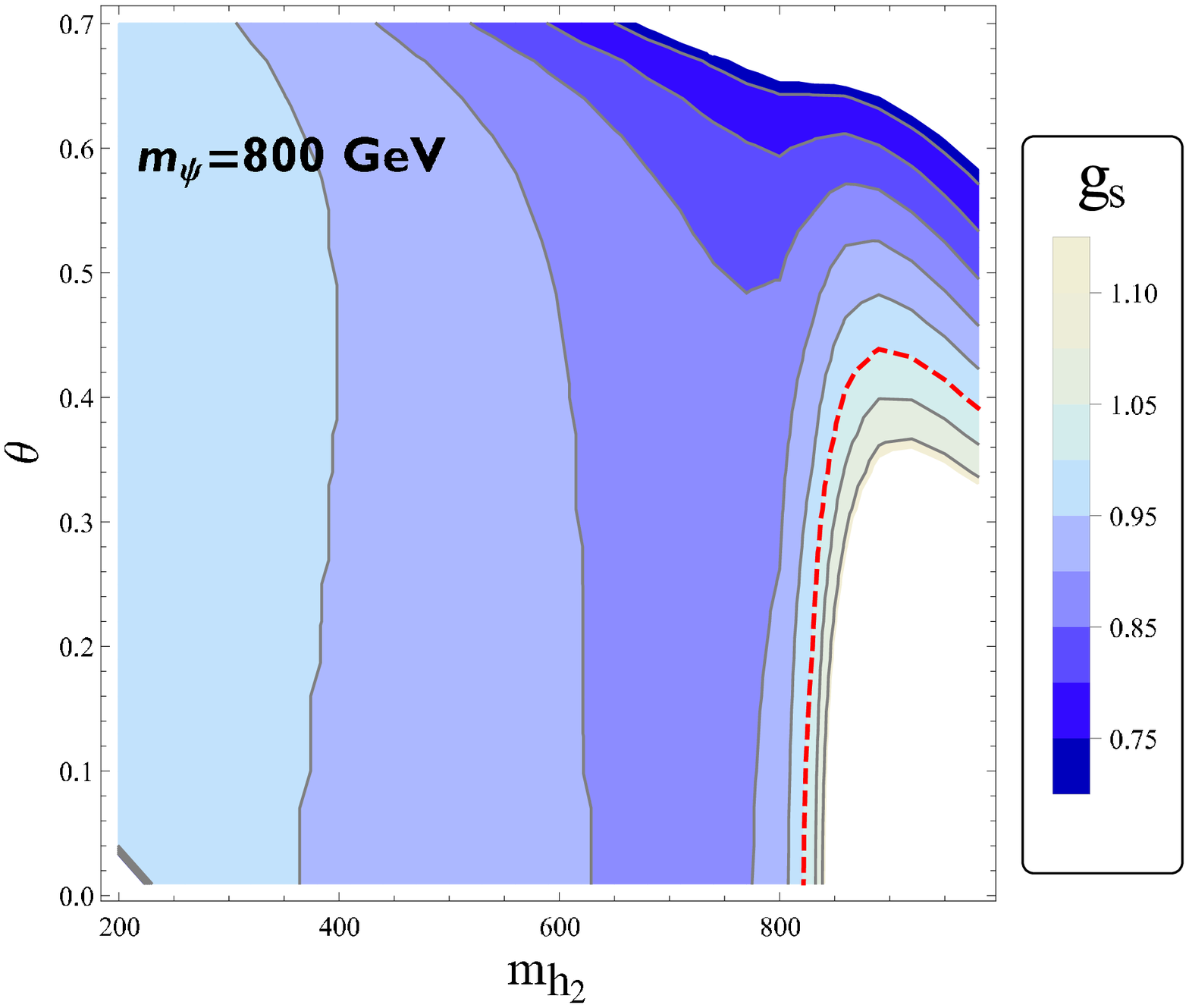}}(b)\raisebox{-2cm}{\includegraphics[width=8cm]{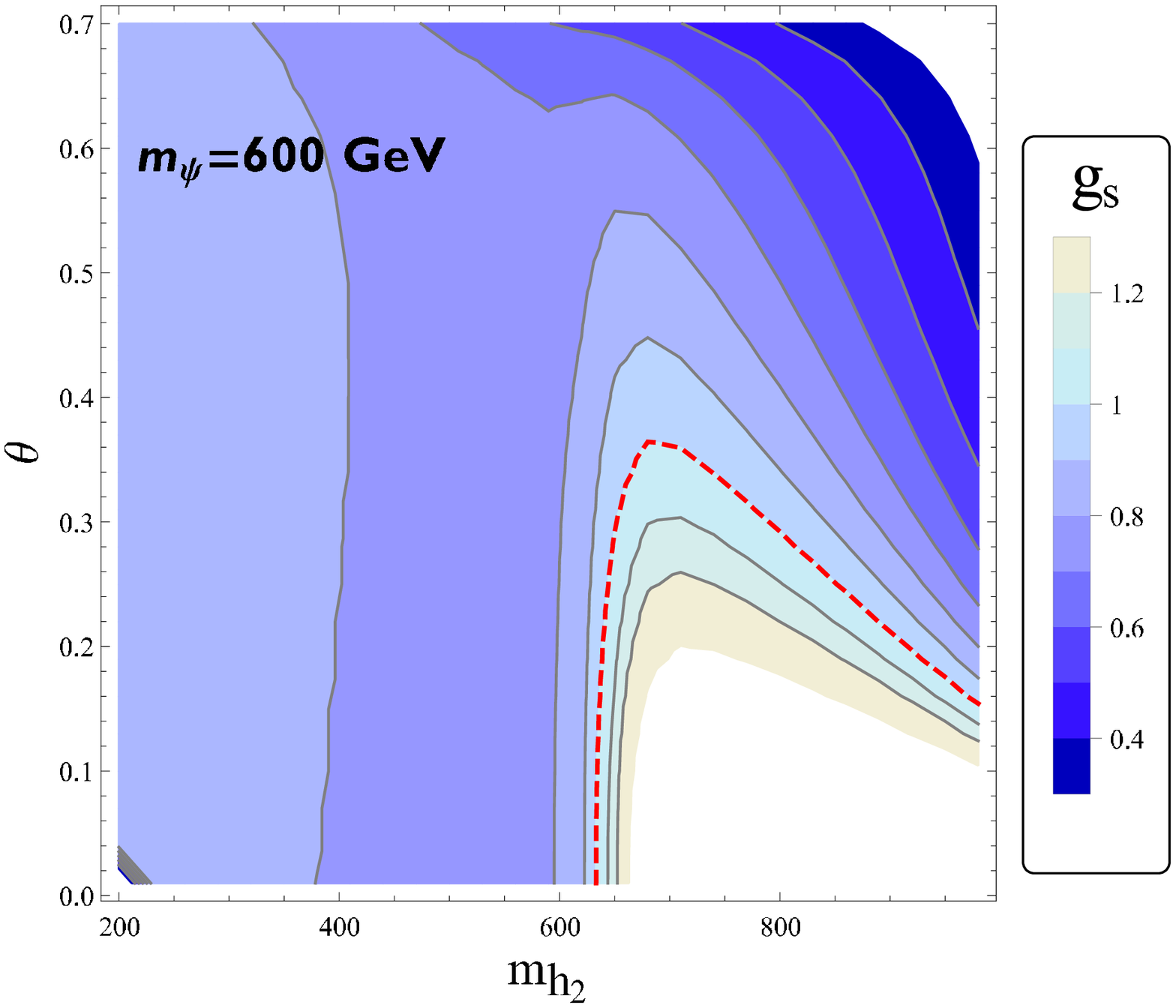}}\\
(c)\raisebox{-2cm}{\includegraphics[width=8cm]{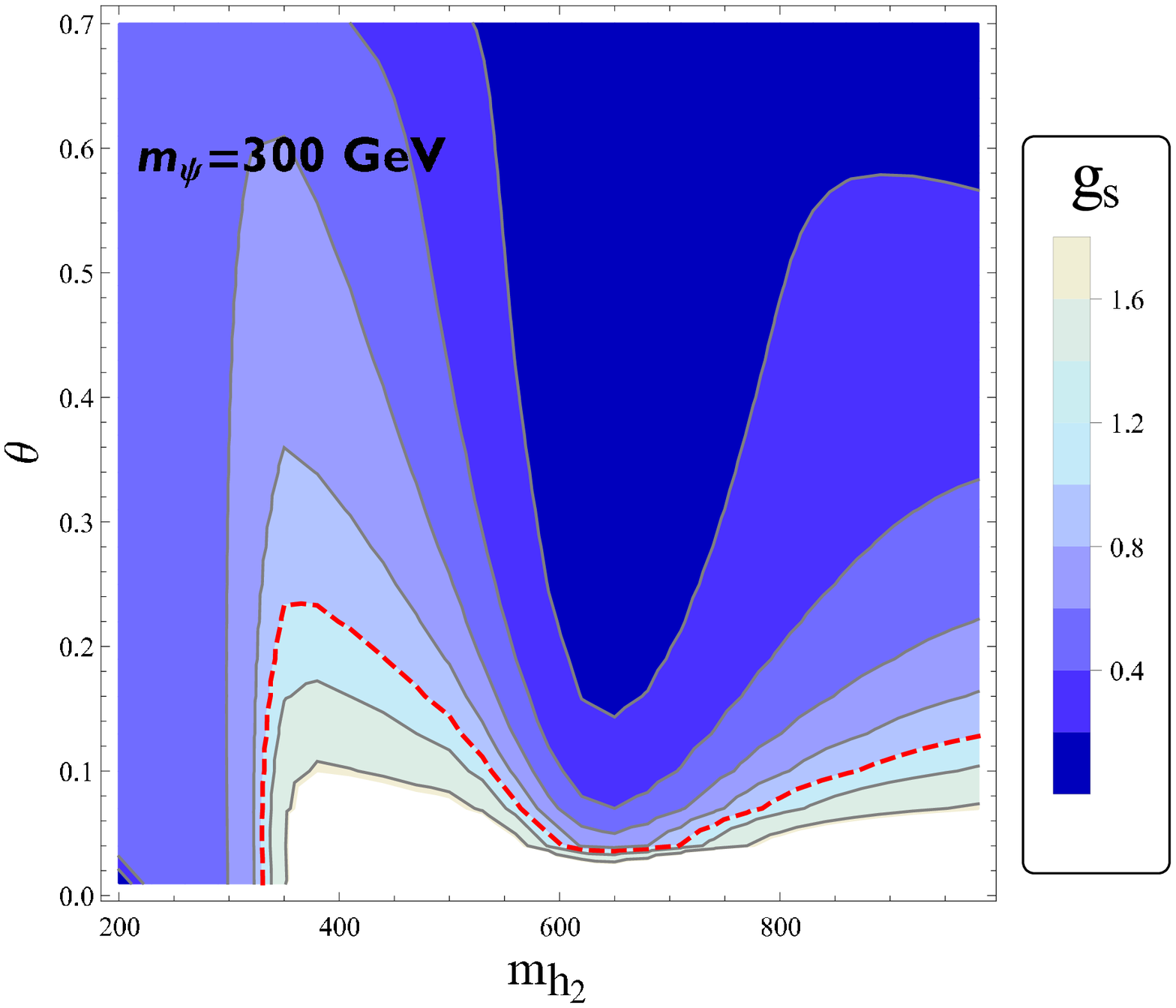}}(d)\raisebox{-2cm}{\includegraphics[width=8cm]{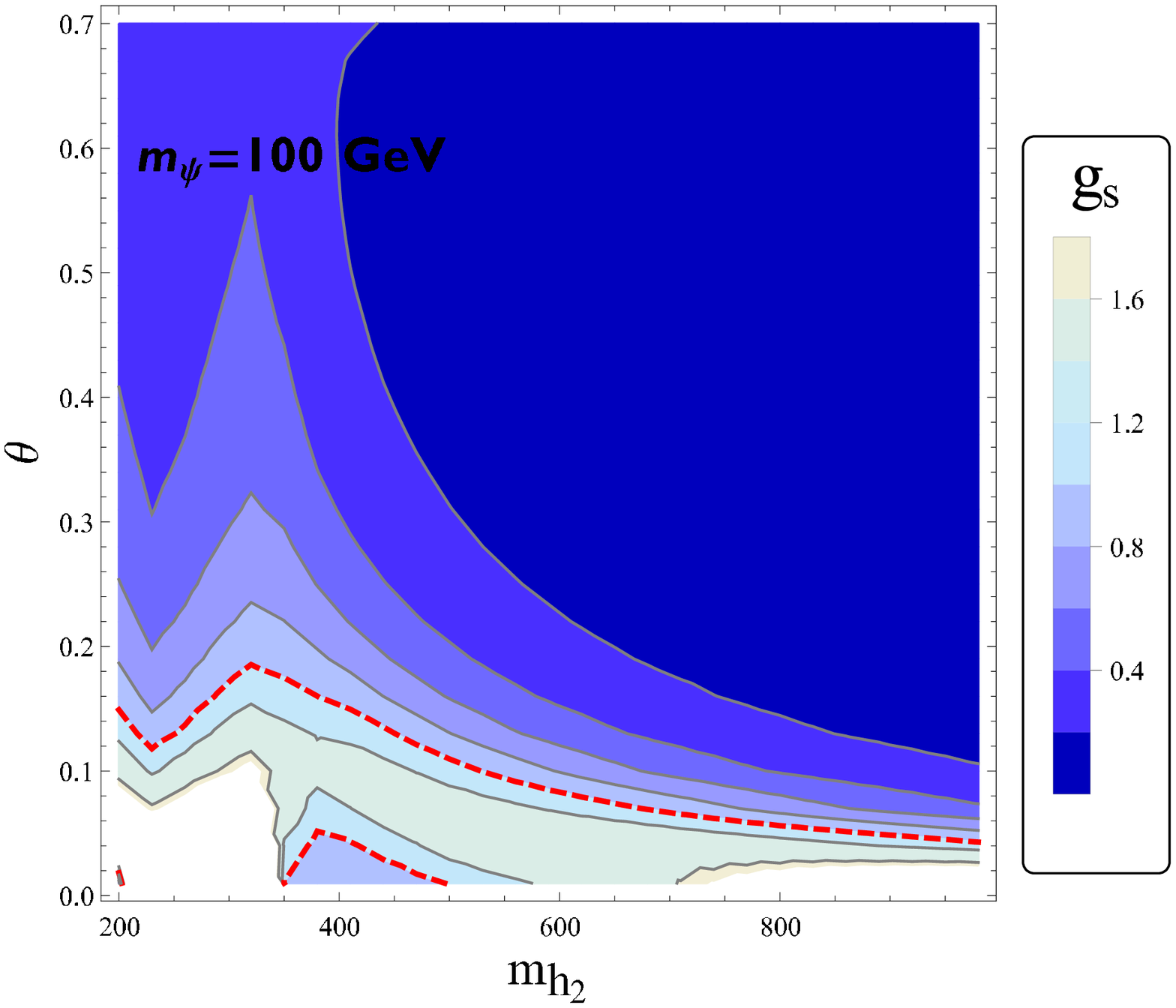}}}
\caption{\small  Contour plots for $g_s$ vs Higgs bosons mixing angel $\theta$, and $m_{h_2}$ corresponds to different values of $m_\psi$. Dashed line shows $g_s=1$.
 }\label{f3}
\end{figure}
In Fig. \ref{f3}, we have given four contourplots which illustrate variations of $g_s$ in terms of the mass of the second Higgs boson and the mixing angle for various SFCDM mass. Via this figures, one can explore parts of parameter space to find regions in which perturbation theory and WMAP constraint are both satisfied.

\begin{figure}[th]
\centerline{\includegraphics[width=12.5cm]{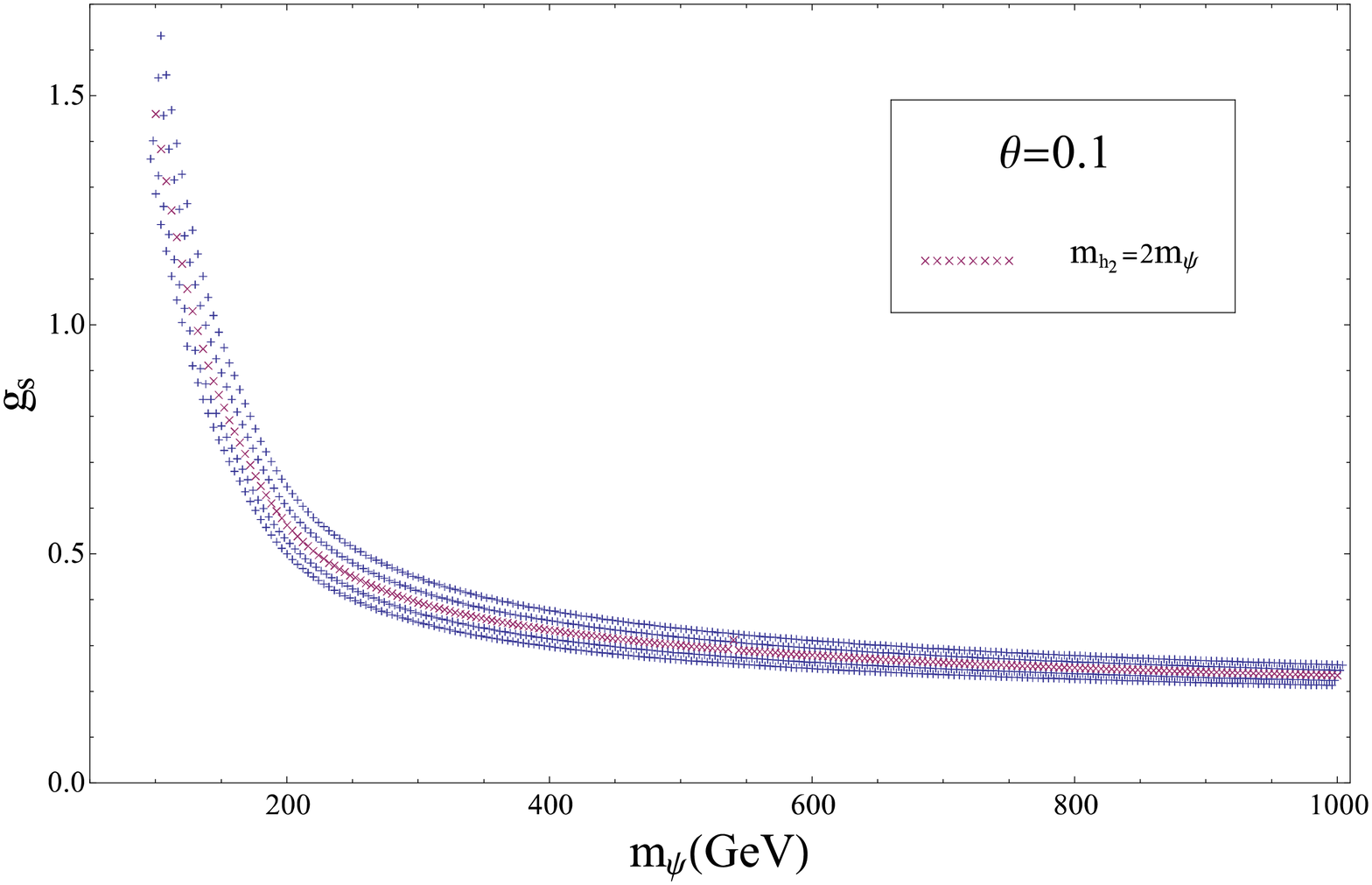}}
\caption{\small  $g_s$ vs SFCDM mass at resonance region for $\theta=0.1$. For $m_\psi$ larger than about 150 GeV we can use $g_s$ for our direct detection calculation.
 }\label{f4}
\end{figure}

Before we discussed the direct detection constraints, we should  notice that the perturbation theory used here is applicable for $\theta=0.7$. For $\theta=0.1$ at resonance ($m_\psi\sim m_{h_2}/2$) we can sure that our calculations work properly.  Fig. \ref{f4} shows the variation of $g_s$ vs SFCDM mass about resonance, i.e. when the SFCDM mass is about half of the second Higgs mass.

\section{Direct detection constraints}
\label{s3}
\begin{figure}[th]
\centerline{\includegraphics[width=12.5cm]{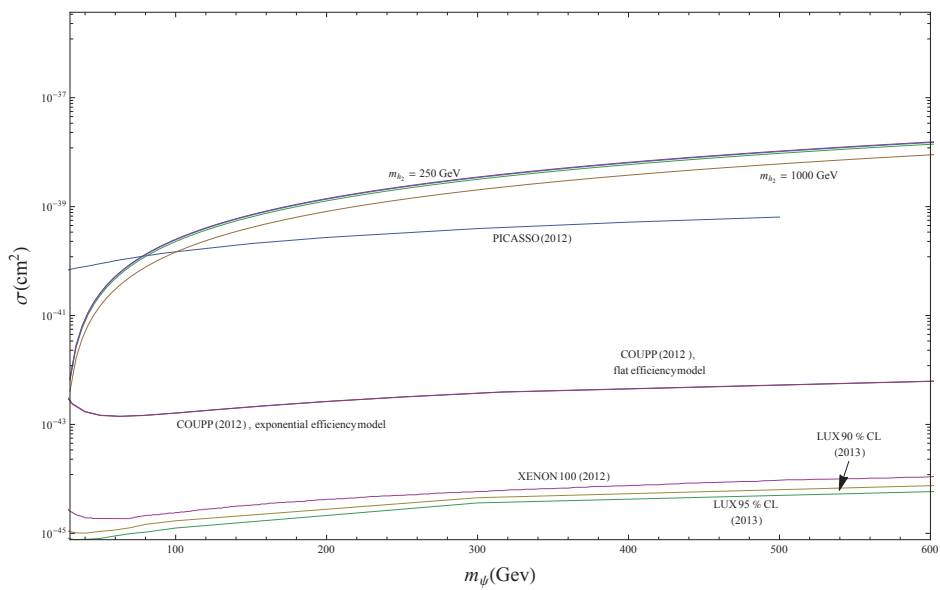}}
\caption{\small  Scattering cross section of SFCDM  off nucleons vs its mass. Four different  values for $m_{h_2}$ is selected and $\theta=0.7$. The theoretical results are compared with the recent existing experimental data.}\label{f5}
\end{figure}
\begin{figure}[th]
\centerline{\includegraphics[width=12.5cm]{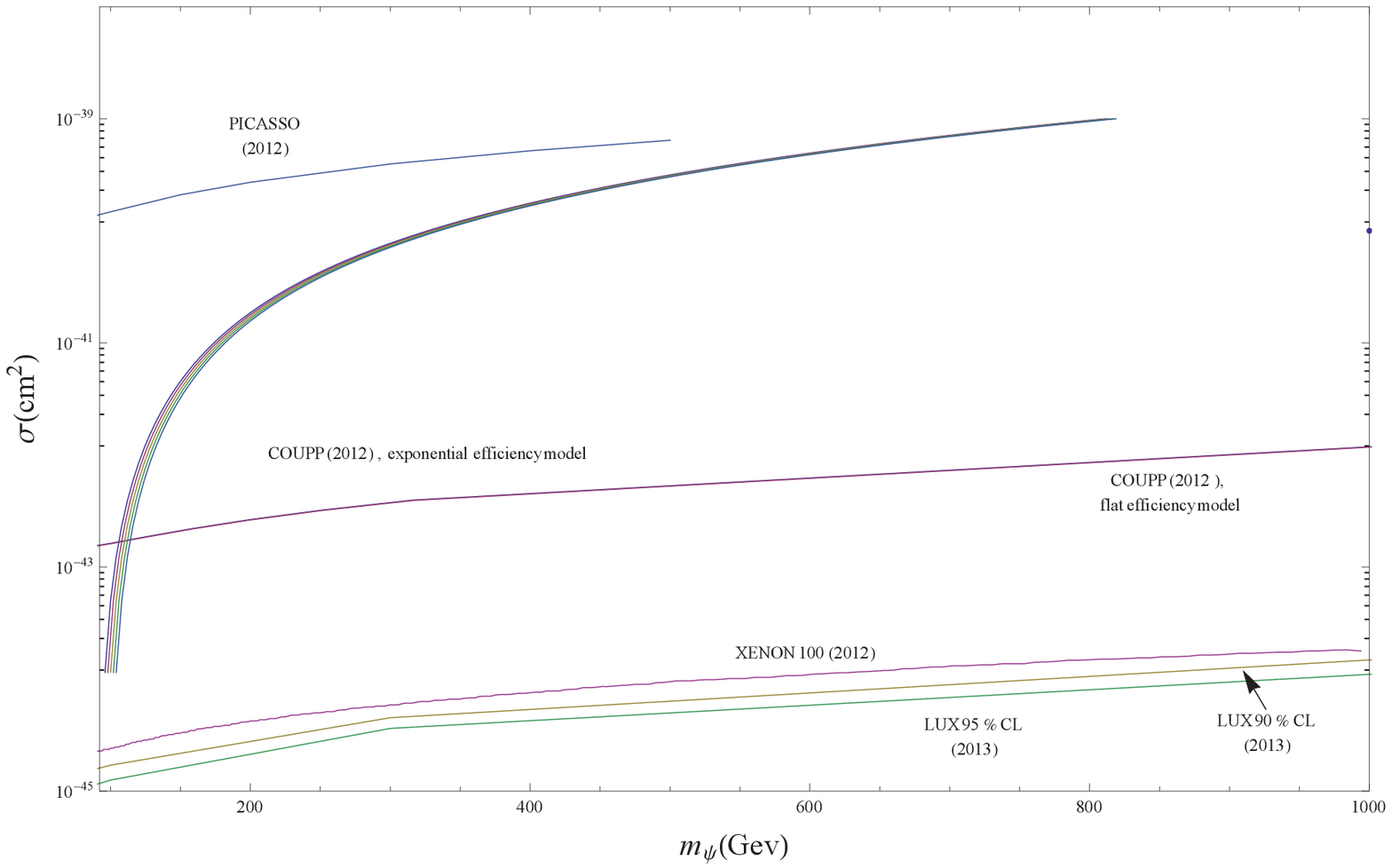}}
\caption{\small   Scattering cross section of SFCDM  off nucleons vs its mass. Here $m_{h_2}$ is about twice of dark matter mass (resonance region) and $\theta=0.1$. The theoretical results are compared with the recent existing experimental data.
 }\label{f6}
\end{figure}
In addition to the relic abundance, one can constrain the parameter space of a DM theory by the direct detection bounds. In this section, we explore the parameter space of the SFCDM model which is consistent with relic abundance constraint by direct detection data. The elastic spin-independent cross section of the scattering of SFCDM from a nucleon is described by the following effective Lagrangian at the hadronic level:
\begin{equation}
\label{eff}
  {\cal L}_{\text{eff}} =f_{\text{p}}(\bar{\psi}\psi)(\bar{\text{p}}\text{p})+f_{\text{n}}(\bar{\psi}\psi)(\bar{\text{n}}\text{n}),
\end{equation}
where $f_{\text{p}}$ and $f_{\text{n}}$  are respectively the effective couplings of DM to protons and neutrons, and given by:
\begin{equation}\label{6}
 \frac{ f_{\text{p,n}}}{m_{\text{p,n}}}=\sum_{q=u,d,s} f_{Tq}^{\text{(p,n)}}\frac{\alpha_{q}}{m_{q}}+\frac{2}{27}f_{Tg}^{\text{(p,n)}}\sum_{q=c,b,t} \frac{\alpha_{q}}{m_{q}},
\end{equation}
with the matrix elements $m_{\text{p,n}}f_{Tq}^{\text{(p,n)}}\equiv \langle \text{p,n}|m_{q}\bar{q}q|\text{p,n}\rangle$  for $q=u,d,s$ and $f_{Tg}^{\text{(p,n)}} =1-\sum\limits_{q=u,d,s} f_{Tq}^{\text{(p,n)}}$. The numerical values of the hadronic matrix elements are given in \cite{Ellis} (to see improved theory predictions for the coupling of the scalar quark current to the nucleon relevant for the direct detection cross section refer to \cite{Crivellin})
\begin{equation}
 f_{ Tu}^{\text{(p)}} = 0.020\pm0.004 \hspace{5mm};\hspace{5mm} f _{Td}^{\text{(p)}}=0.026\pm0.005 \hspace{5mm};\hspace{5mm} f_{ Ts}^{\text{(p)}} = 0.118\pm0.062,
\end{equation}
and
\begin{equation}
 f_{ Tu}^{\text{(n)}} = 0.014\pm0.003 \hspace{5mm};\hspace{5mm} f _{Td}^{\text{(n)}}=0.036\pm0.008 \hspace{5mm};\hspace{5mm} f_{ Ts}^{\text{(n)}} = 0.118\pm0.062.
\end{equation}
Since $f_{ Ts}^{\text{(p)}}$ and $f_{ Ts}^{\text{(n)}}$ are equal and have the dominate contribution in $f_{\text{p}}$ and $f_{\text{n}}$, we let $f_{\text{p}}\approx f_{\text{n}}$. Here, $\alpha_q$ is an effective coupling constant between SFCDM and quark $q$, according to the following effective Lagrangian:
\begin{equation}
{\cal L}_{\text{eff}}=\sum_q\alpha_q\bar{\psi}\psi\bar{q}q.
\end{equation}
The scattering SFCDM and quarks proceeds through $t$-channel diagram with intermediating Higgs bosons and $\alpha_q$ is, consequently, determined as follows:
\begin{equation}
\alpha_q=\frac{g_s \sin\theta\cos\theta m_q}{v_0}\left(\frac{1}{m_{h_1}^2}-\frac{1}{m_{h_2}^2}\right).
\end{equation}
Therefore, the elastic scattering cross section with a single nucleon is given by
\begin{equation}\label{j}
 \sigma(\psi \text{p}\rightarrow \psi \text{p})= \frac{4m_\text{r}^{2}}{\pi}f_{\text{p}}^{2},
\end{equation}
where $m_\text{r} =\left(\frac{1}{m_{\psi}}+\frac{1}{m_{\text{p}}}\right)^{-1}$. We figure out the cross section $\sigma$ for $\theta=0.7$ (for such a mixing angle $g_s$ is less than one almost everywhere) and various values of $m_{h_2}$ in Fig. \ref{f5}. In this figure, we compare our theoretical results with recently published experimental data of XENON100 \cite{XENON100}, LUX \cite{LUX}, COUPP \cite{coupp} and PICASSO \cite{picasso} collaborations.  As we see, this theory is excluded by all the mentioned experimental bounds for SFCDM masses larger than 100 GeV and maximal Higgs mixing. As we discussed in the previous section, the perturbation theory is destroyed when the Higgs mixing is non-maximal unless at resonances. Hence, we study the direct detection for $\theta=0.1$ and $m_\psi\sim\frac{1}{2}m_{h_2}$ via Fig. \ref{f6}. In this figure, we see that the model is also excluded by the mentioned direct detection experiments except for $m_{h_2}\sim 2m_\psi<200$ GeV where is excluded by LHC.
\section{Summary and discussion}
As a minimal and renormalizable theory for explanation of CDM, one can extend SM by a singlet fermion as CDM and a singlet Higgs as the mediator between SFCDM and SM particles \cite{sf2}. This theory has seven parameters in addition to the SM ones: singlet fermion mass $m_\psi$, a coupling constant between singlet fermion and singlet Higgs $g_S$, the mass of the singlet Higgs $m_0$ and its self-interaction couplings $\lambda_3$ and $\lambda_4$, and the coupling constants between singlet Higgs and SM one $\lambda_1$ and $\lambda_2$. After electroweak symmetry breaking, the new set of parameters consists of: $m_\psi$, $g_S$, mixing angel between Higgs bosons $\theta$, second Higgs mass $m_{h_2}$, $\lambda_1$, $\lambda_2$, $\lambda_3$ and $\lambda_4$ (one of which is not an independent parameter). We have computed the complete annihilation cross section of singlet fermion pair into the SM particles and new Higgs boson at tree level in perturbation theory. We have investigated the parameter space under relic abundance constraint for dark matter masses up to 1 TeV,  independent of \cite{recent}. Although in this reference authors try to analysis SFCDM model based on a sample of about $10^5$ random models, the role of each parameter is not clearly specified. In addition, it is clear that the perturbation condition is not respected in their work. However, in the present paper the effect of each parameter has been investigated separately and we work in a self-consistent way with perturbation theory. We find that $\lambda_1$, $\lambda_2$ and $\lambda_3$ do not play an important role (please see Fig. \ref{f1} for $\lambda_2$). The cross section dependencies on $g_s$, $\theta$ and $m_{h_2}$ have been explored through Figs. \ref{f2} and \ref{f3}. As we see, the maximal mixing of Higgs bosons leads to $g_s<1$ which is required for perturbation theory. For non-minimal Higgs bosons mixing, $g_s$ is larger or about one except usually at resonances. Fig. \ref{f4} illustrates $g_s$ vs $m_\psi$ for $m_\psi\sim m_{h_2}/2$ and $\theta=0.1$,  for instance. We have also studied the direct detection bounds in Section \ref{s3}. We obtained scattering cross section of SFCDM off nucleons for almost maximal Higgs bosons mixing, $\theta=0.7$, (Fig. \ref{f5}) and a minimal Higgs bosons mixing, $\theta=0.1$, at resonance (Fig. \ref{f6}). We have compared our results with experimental ones reported by XENON100 \cite{XENON100}, LUX \cite{LUX}, COUPP \cite{coupp}  and  PICASSO \cite{picasso} collaborations. It is clear from these figures that the SFCDM is excluded by these experiments for choosing parameters which are consistent with perturbation theory and relic abundance constraints. Of course, one should note that there exists another process through which a density of the singlet fermion can be produced;  DM can be treated as a feebly interacting massive particle (FIMP) which consider  in \cite{hall}. A FIMP can be produced trough `freeze-in' mechanism. In this scenario, a particle has been no longer in equilibrium and its density varies from zero (at very high temperatures in early universe)  to a constant value. Model independently,  the coupling constant of DM to the SM particles is of order of $10^{-11}$. Therefore, even though the whole of the parametric space is excluded experimentally, the FIMP scenario can remain as an alternative mechanism.

\vskip20pt\noindent {\large {\bf
Acknowledgements}}\vskip5pt\noindent

It is a great pleasure for us to acknowledge the useful discussion and comments
of S.M. Fazeli.

\section{Annihilation cross section}
In this appendix we calculate the annihilation cross section of singlet fermion into the other particles accommodated in our theory at tree level. While the annihilations into the fermions and gauge bosons proceed through $s$-channel, the annihilation into Higgs bosons occurs via $s$-, $t$- and $u$-channels (see figure \ref{fig1}. The total annihilation cross section can be written as follows:
\begin{equation}\label{sigmaa}
 { {\sigma v}}_{\text{ann}}  = {\sigma v}_{\text{SM}}+{\sigma v}_{\text{2Higgs}}+ {\sigma v}_{\text{3Higgs}},
\end{equation}
where the $ { {\sigma v}}_{\text{SM}} $ is given by
\begin{eqnarray}
\sigma v_{\text{SM}} &=& \frac{(g_s s_1 s_2)^2}
                        {16\pi}
                   \left(1-\frac{4 m_{\psi}^2}{{s}}\right)
\nonumber\\
      & & \times \left(\sum_{j=1,2}\frac{1}
            {d_j}
            -\frac{2({s}-m_{h_1}^2)({s}-m_{h_2}^2)
                    + 2 m_{h_1}m_{h_2}\Gamma_{h_1}\Gamma_{h_2}}
                  {d_1d_2}
      \right)
\nonumber\\
      & & \times \left[\sum_{f(\text{fermions})}
           2\lambda_fs\left(\frac{m_f}{v_0} \right)^2 \left(1-\frac{4 m_f^2}{{s}}\right)^{3/2}
      \right.
\nonumber\\
      & & ~~~~~
         \left.  + \sum_{{w=W^{^+},W^{^-},Z^0}}2\left(\frac{m_w^2}{v_0}\right)^2
             \left(2+\frac{({s}-2m_w^2)^2}{4 m_w^4}\right)
                 \sqrt{1-\frac{4 m_w^2}{{s}}}\right],
\end{eqnarray}
where $\lambda_f$ is three (one) for quarks (leptons), $\Gamma_{h_i}$ refers to the decay widths of $h_i$ and $d_j=({s}-m_{h_j}^2)^2 + m_{h_j}^2 \Gamma_{h_j}^2$. Here, we have used these abbreviations: $s_1\equiv\sin\theta$ and $s_2\equiv\cos\theta$. The last two terms in Eq. (\ref{sigmaa}), are  the annihilation cross sections into two and three Higgs bosons, respectively. To obtain these cross sections we should derive $g_{jkl}$ and $g_{jklm}$ corresponding to the vertex factors of three and four Higgs boson lines, respectively. For $j\neq k$ we get
\begin{eqnarray}\label{couplings}
g_{{jjj}}&=&\frac{1}{3} \left\{6 (-1)^j v_0 s_k \left(\lambda _2 s_j^2+\lambda _0 s_k^2\right)-s_j \left[s_j^2 \left(\lambda _3+\lambda _4 x_0\right)+3 \lambda _1 s_k^2\right]\right\}
   \nonumber\\
g_{{jkk}}&=&\frac{1}{4} \left\{2 (-1)^k v_0 s_j \left[\lambda _2 \left(1-3 s_j^2\right)+3 \left(4 \lambda _0-3 \lambda _2\right) s_k^2\right]
\right.
 \nonumber\\
&& \qquad\left.
+s_j^2 s_k \left[9 \lambda _1-4 \lambda _3+2 \left(9 \lambda _2-2 \lambda _4\right) x_0\right]-\left(3 s_k^3+s_k\right) \left(\lambda _1+2 \lambda _2 x_0\right)\right\}
\nonumber\\
g_{jjjj}&=&-12 \lambda _2 s_1^2s_2^2-\lambda _4 s_j^4-6 \lambda_0 s_k^4
 \nonumber\\
 g_{1122}&=&\frac{1}{8}\left\{[\cos (4 \theta )-1] \left(\lambda _4+6 \lambda_0\right)-4 \lambda _2 [3 \cos (4 \theta )+1]\right\}
\nonumber\\
g_{jjjk}&=&s_2 s_1\left(6 \lambda _2 (s_j^2-s_k^2)-\lambda _4 s_j^2+6 \lambda_0 s_k^2\right).
\end{eqnarray}
Note that $g_{ijk}$ and $g_{ijkl}$ are symmetric under permutation of their subscripts. Therefore one can derive the annihilation cross section into two Higgs bosons as follows:
\begin{eqnarray}
\sigma v_{\text{2Higgs}}&=&\frac{g_s^2}{16 \pi } \left(1-\frac{4 m_{\psi }^2}{s}\right) \left\{-\frac{4 g_s^2 s_1^2 s_2^2}{y \left(y^2-1\right) \left(-m_{h_1}^2-m_{h_2}^2+s\right){}^2}
\right.
\nonumber\\&\times&
\left.
 \left\{\left(-m_{h_1}^2-m_{h_2}^2+s\right){}^2 y^3+\left[-32 m_{\psi }^4+8 \left(m_{h_1}^2+m_{h_2}^2\right) m_{\psi }^2-m_{h_1}^4-\left(m_{h_2}^2-s\right){}^2-m_{h_1}^2 \left(4 m_{h_2}^2-2 s\right)\right] y
\right.\right.
\nonumber\\&&
\left.\left.
+\left(y^2-1\right) \tanh ^{-1}y \left[32 m_{\psi }^4+8 \left(m_{h_1}^2+m_{h_2}^2-2 s\right) m_{\psi }^2-m_{h_1}^4-\left(m_{h_2}^2-s\right){}^2+2 m_{h_1}^2 \left(s-2 m_{h_2}^2\right)\right]\right\}
\right.
\nonumber\\&-&
\left.
\frac{8g_s m_{\psi } s_1 s_2}{d_1 d_2} \left[\frac{\tanh ^{-1}y \left(8 m_{\psi }^2-m_{h_1}^2-m_{h_2}^2-s\right)}{y \left(-m_{h_1}^2-m_{h_2}^2+s\right)}-1\right] \left[d_2 g_{112} \left(s-m_{h_1}^2\right) s_1+d_1 g_{212} \left(s-m_{h_2}^2\right) s_2\right]
\right.
\nonumber\\&+&
\left.
\sqrt{{\frac{\left(-m_{h_1}^2-m_{h_2}^2+s\right){}^2-4 m_{h_1}^2 m_{h_2}^2}{s^2}}} \left[\frac{2 g_{112} g_{212} s_1 s_2}{d_1 d_2} \left[\left(s-m_{h_1}^2\right) \left(s-m_{h_2}^2\right)+m_{h_1} m_{h_2} \Gamma _{h_1} \Gamma _{h_2}\right]+\sum _{j=1,2} \frac{g_{{j12}}^2 s_j}{d_j}\right]
\right.
\nonumber\\&+&
\left.
\frac{1}{2} \sum _{k=1,2} \left[\frac{g_s^2  s_k^4}{ x_k \left(x_k^2-1\right)\left(s-2 m_{h_k}^2\right)^2} \left[4 x_k \left(32 m_{\psi }^4-16 m_{h_k}^2 m_{\psi }^2+6 m_{h_k}^4+s^2-4 s m_{h_k}^2-\left(s-2 m_{h_k}^2\right){}^2 x_k^2\right)
\right.\right.\right.
\nonumber\\&&
\left.\left.\left.
-4\left(x_k^2-1\right) \tanh ^{-1}x_k \left(32 m_{\psi }^4+16 \left(m_{h_k}^2-s\right) m_{\psi }^2-6 m_{h_k}^4-s^2+4 s m_{h_k}^2\right) \right]
\right.\right.
\nonumber\\&&
\left.\left.
+\sqrt{1-\frac{4 m_{h_k}^2}{s}} \left(\frac{2 g_{{1kk}} g_{2 {kk}} s_1 s_2}{d_1 d_2}{ \left[\left(s-m_{h_1}^2\right) \left(s-m_{h_2}^2\right)+m_{h_1} m_{h_2} \Gamma _{h_1} \Gamma _{h_2}\right]}+\sum _{j=1,2} \frac{g_{{jkk}}^2 s_j}{d_j}\right)
\right.\right.
\nonumber\\&&
\left.\left.
-\frac{8 \left({g_s} m_{\psi } s_k^2\right)}{d_1 d_2}{ \left(\frac{\tanh ^{-1}x_k\left(-8 m_{\psi }^2+2 m_{h_k}^2+s\right)}{\left(2 m_{h_k}^2-s\right) x_k}-1\right) \sum _{j=1,2} g_{{jkk}} \left(s-m_{h_j}^2\right) s_j d_j}\right]\right\}
\end{eqnarray}
where we have
\[x_k={\sqrt{1-\frac{4 m_{\psi }^2}{s}} \sqrt{1-\frac{4 m_{h_k}^2}{s}}}\bigg/{\left(1-\frac{2 m_{h_k}^2}{s}\right)}\]
 and
 \[y=-{\sqrt{1-\frac{4 m_{\psi}^2}{s}} \sqrt{\frac{m_{h_1}^4}{s^2}+\left(\frac{m_{h_2}^2}{s}+1\right) \left(1-\frac{2 m_{h_1}^2+m_{h_2}^2}{s}\right)}}\bigg/{\left(1-\frac{m_{h_1}^2+m_{h_2}^2}{s}\right)}.\]
Although the annihilation cross section into three Higgs bosons suppressed due to its narrow phase space integral, to have a complete and precise  calculation we take it into account. For this term we have
\begin{eqnarray}
\sigma v_{\text{3Higgs}} =\frac{2g_{s}^{2}(s-4 m_{\psi}^{2})}{1536\pi^{3}}\sum _{k,l,m}\left(\sum _{j=1,2} \frac{g_{{jklm}}^2 s_j^2}{d_j}+\frac{2 s_1 s_2 g_{1{klm}} g_{2{klm}} \left[\Gamma _{h_1} \Gamma _{h_2} m_{h_1} m_{h_2}+\left(s-m_{h_1}^2\right) \left(s-m_{h_2}^2\right)\right]}{d_1 d_2} \right).
\end{eqnarray}

\end{document}